\documentclass[fleqn,usenatbib]{mnras}

\usepackage[T1]{fontenc}
\usepackage{ pdflscape}
\DeclareRobustCommand{\VAN}[3]{#2}
\let\VANthebibliography\thebibliography
\def\thebibliography{\DeclareRobustCommand{\VAN}[3]{##3}\VANthebibliography}

\usepackage{graphicx}	
\usepackage{amsmath}	
\usepackage{amssymb}	
\usepackage{multirow}
\usepackage{booktabs}
\usepackage{diagbox}
\usepackage[english]{babel}

\newcommand{\teff}{\ensuremath{T_\mathrm{eff}}}

\newcommand{\logg}{\ensuremath{\log g}}
\newcommand{\kms}{$\rm km\,s^{-1}$}
\newcommand{\ms}{$\rm m\,s^{-1}$}

\newcommand{\kmsmath}{\rm km\,s^{-1}}

\newcommand{\thisplain}{\texttt{emcee}}

\title[]{Accurate mass-radius ratios for Hyades white dwarfs}

\author[L. Pasquini et al.]{
L. Pasquini,$^{1}$\thanks{E-mail: lpasquin@eso.org, Observations based on ESO observing program 0106.D-0972(A) and 0108.D-0872(A) }
A.~F. Pala,$^{2}$
M. Salaris,$^{3,4}$
H.-G. Ludwig, $^{5}$
I. Le\~ao,$^{6}$
A. Weiss,$^{7}$
and 
J.~R. de Medeiros$^{6}$
\\
$^{1}$  ESO - European Southern Observatory, Karl-Schwarzchild-Strasse 2, 85748 Garching bei M\"{u}nchen, Germany \\
$^{2}$  European Space Agency, European Space Astronomy Centre, Camino Bajo
del Castillo s/n, Villanueva de la Ca\~nada, E-28692 Madrid, Spain\\
$^{3}$  Astrophysics Research Institute, Liverpool John Moores University, 146 Brownlow Hill, Liverpool L3 5RF, UK \\
$^4$ INAF  Osservatorio Astronomico di Abruzzo, Via M. Maggini, s/n, I-64100 Teramo, Italy \\
$^5$ Zentrum f\"{u}r Astronomie der Universit\"at Heidelberg. Landessternwarte, K\"onigstuhl 12, 69117 Heidelberg, Germany \\ 
$^{6}$  Departamento de Fisica, Universidade Federal do Rio Grande do Norte, 59078-970 Natal, RN, Brazil\\
$^{7}$  Max Plank f\"{u}r Astrophysiik, Karls Schwarzschild Str. 1, 85748 Garching bei M\"{u}nchen, Germany\\
}

\date{Accepted XXX. Received YYY; in original form ZZZ}

\pubyear{2022}

\begin{document}
\label{firstpage}
\pagerange{\pageref{firstpage}--\pageref{lastpage}}
\maketitle

\begin{abstract}
We use the ESPRESSO spectrograph at the Very Large Telescope to measure velocity shifts and gravitational redshifts of eight bona fide Hyades white dwarfs, with an accuracy better than 1.5 percent. By comparing the gravitational redshift measurements of the mass-to-radius ratio with the same ratios derived by fitting the \textit{Gaia} photometry with theoretical models, we find an agreement to better than one per cent. It is possible to reproduce the observed white dwarf cooling sequence and the trend of the mass-to-radius ratios as a function of colour using isochrones with ages between 725 and 800 Myr, tuned for the Hyades. One star, EGGR\,29, consistently stands out in all diagrams, indicating that it is possibly the remnant of a blue straggler. We also computed mass-to-radius ratios from published gravities and masses, determined from spectroscopy. The comparison between photometric and spectroscopic stellar parameters reveals that spectroscopic effective temperature and gravity are systematically larger than the photometric values. Spectroscopic mass-to-radius ratios disagree with those measured from gravitational redshift, indicating the presence of systematics affecting the white dwarf parameters derived from the spectroscopic analysis. 
     
\end{abstract}

\begin{keywords}
Stars: white dwarfs -- Stars: masses -- Stars: radii-- Stars: Open Clusters --Techniques: spectroscopy-- ESPRESSO
\end{keywords}



\section{Introduction}

White dwarfs (WDs) are the most common stellar remnant in the Galaxy, being the final evolutionary stage of low- and intermediate-mass stars ($M\!\lesssim\!8-10\,\mathrm{M}_\odot$, \citealt{Siess2007}), which represent over $\simeq 95\,$per cent of the stars in the Milky Way \citep[e.g.][]{Althaus+2010}.
The accurate determination of their physical parameters is relevant to address several astrophysics questions. First, thanks to their long cooling times, WDs are a powerful tool to estimate the age of the Galactic disc \citep{Fantin+2019} and halo \citep{Kalirai2012}, and to investigate the local star formation history and the initial mass function.
Moreover, the difference between the WD masses and the initial masses of their progenitors provides the initial-final mass relation for single stars \citep[IFMR -- see, e.g.][]{Kalirai+2008, Salaris+2009} that is essential to model stellar populations and, more generally, to verify predictions of stellar evolutionary models. Finally, the IFMR can be used to test the predicted integrated mass loss of low- and intermediate-mass stars, a key ingredient in the chemical evolution of galaxies. 

It is therefore important to obtain accurate and reliable measurements of WD masses, that can be derived employing various methods.
One possibility is to employ WD colour-magnitude diagrams (CMDs): the interpolation of grids of WD cooling models to the observed colours and absolute magnitudes provides the mass (and cooling time) of the targets. For example, using \textit{Gaia} photometry and parallaxes for WDs in the Hyades cluster, \citet{Salaris+2018} determined masses with a precision of $1-2$ per cent.
Alternatively, it is possible to determine the WD surface gravity (\logg) and the effective temperature (\teff) by employing synthetic spectra to fit the observed Balmer lines \citep[see, e.g.,][]{2011ApJ...743..138G} or spectral energy distribution \citep[e.g.][]{Raddi+2017}. If the distance to the system is known and assuming a mass-radius relationship, it is possible to derive WD masses with a precision of a few percent. 

There are also other methods that minimise the input from theoretical models. One is the full solution of binary systems hosting a WD, including  eclipsing binaries, that can accurately be applied to only a few systems \citep[e.g][]{Joyce+2018a,Parsons+2017,2022AJ....163...34M}. Another method is the comparison between the WD astrometric radial velocity and the spectroscopic velocity shift along the line of sight, $\Delta V$, which provides a measurement of the WD gravitational redshift and, thus, a  direct determination of the stellar mass-to-radius ratio ($M/R$) \citep[e.g.][]{Luca+2019}.
Here we use this latter method to determine accurate M/R for WDs in the Hyades open cluster.

The Hyades cluster is the ideal laboratory for this study. First, given its proximity, the parallax provided by \textit{Gaia} is accurate to $\simeq\!0.2$ per cent (the centre-of-mass distance is equal to $47.50 \pm 0.15\,$pc, \citealt{Gaia_Collaboration2018}). Second, the astrometric radial velocities of the Hyades stars have smaller uncertainties than the cluster velocity dispersion, which amounts to $\lesssim\!340\,\mathrm{m\,s}^{-1}$ \citep{Leao+2019}. Consequently, spectral $\Delta V$ 
measurements of the WDs in the Hyades allow us to determine the gravitational redshift with an uncertainty of $\simeq\!340\,\mathrm{m s}^{-1}$, which yields an accuracy better than one percent for $M/R$.

\cite{Luca+2019} determined the masses and $M/R$ ratios of six Hyades WDs from gravitational redshift measurements, by combining UVES archive spectra obtained by the Supernovae Type Ia Progenitor Survey (SPY) \citep{Napiwotzki+2001} with published results from the analysis of data from the HIRES spectrograph at Keck. The masses derived for all targets were systematically lower than those derived with other methods, such as the spectroscopic fit of the Balmer lines and the analysis of the photometric spectral energy distribution. The uncertainty in the $\Delta V$ measured by those authors were $\simeq\,2\,\mathrm{km\,s}^{-1}$, which translates into an uncertainty of $\simeq\!5-10$ per cent in the derived masses. This fact did not allow \cite{Luca+2019} to reach definitive conclusions about the observed mass differences, although the systematic trend suggests that the discrepancy is real.

The present paper aims at measuring accurate masses and $M/R$ ratios of Hyades WDs from the gravitational redshift using ESPRESSO (Echelle SPectrograph for Rocky Exoplanets and Stable Spectroscopic Observations) at the Very Large Telescope (VLT). 
We use this new spectral data to determine $\Delta V$ values by using the narrow non-local thermodynamic equilibrium (NLTE) cores at the centre of the Balmer lines (see e.g. \citealt{Reid1996,Zuckerman+2013,Luca+2019}) with an accuracy of a few hundreds $\mathrm{m\,s}^{-1}$. This work is organised as follows. Section~2 presents the methods used for measuring the gravitational redshift and for deriving the stellar parameters from  photometry. Section 3 presents the results, while Section~4 summarises the conclusions.


\section{Observations}
The spectra have been acquired during several observing runs during 2020-2022 with ESPRESSO at the VLT \citep{Pepe+2021}. Each star had a minimum of three observations available in the European Southern Observatory (ESO) data archive, from which we retrieved the reduced spectra. 

ESPRESSO is fibre fed and equipped with a double scrambler input fibre that makes the measured $\Delta V$ independent of the centring of the star in the fibre and of seeing. The spectrograph was used in single UT (Unit Telescope) mode, which provides a resolving power of $\sim$140\,000 and a spectral coverage between 3782\,\AA\, and 7887\,\AA. The wavelength calibration accuracy of ESPRESSO reaches up to 24~\ms\, based on peak-to-valley discrepancies between Th-A plus Fabry-Perot and Laser Frequency Comb spectra \citep{Tobias+2021}.  

Provided the WD spectra have a signal-to-noise ratio $S/N \gtrsim 15$, the measurement of the centre of the H$\alpha$ core in a single spectrum is possible with an accuracy better than 2~\kms. 
Three spectra were classified as not usable (class C) by the ESO quality control team because of some instability in ESPRESSO, therefore they were not included in our analysis, as well as those spectra with $S/N <7$. 

\section{Method}
\label{method}

Given that the aim is to determine accurate WD $M/R$ values, it is worth recalling the main ingredients of the measurements and the uncertainties affecting each step. We describe below the theoretical foundation for measuring the gravitational redshift velocity and the methods used for the data analysis.

\subsection{Gravitational redshifts} 
The concept of gravitational redshift ($z_{\rm g}$) was introduced by Einstein, and predicts that the shift in wavelength of electromagnetic radiation in the presence of a gravitational potential $U$ can be approximated by $z_{\rm g} \sim U/c^2$ where $c$ is the speed of light. 
The gravitational redshift velocity $V_{\rm z}$ is defined as follows:
\begin{equation}\label{eq:mr_eq}
V_{\rm z} = 636.3 \frac{M}{\mathrm{M_\odot}} \frac{\mathrm{R_\odot}}{R} ~\mathrm{ms^{-1}}-3.2 ~\mathrm{ms^{-1}}
\end{equation}
where $M$ is the mass of the star and $R$ its radius, and the second term includes the corrections given by the observer distance from the Sun and the Earth's gravitational redshift \citep{prsa+2016}.

$V_{\rm z}$ of a star can be derived from the total stellar velocity shift along the line of sight $\Delta V$, once this has been corrected for (i) the stellar radial velocity ($V_{\rm r}$) and (ii) any additional motions ($V_{\rm add}$) in the stellar atmospheres:
\begin{equation}\label{eqgr}
V_{\rm z} = \Delta V - V_{\rm r} - V_{\rm add}
\end{equation}
%
The radial velocity $V_{\rm r}$ can be written as:
\begin{equation}
V_{\rm r} = V_\mathrm{astro} + V_\mathrm{clus}
\end{equation}
where $V_\mathrm{astro}$ and $V_\mathrm{clus}$ are the astrometric radial velocity and the internal  velocity field of the cluster, respectively.

In the  case of the Hyades, \citet{Leao+2019} have shown that, for non-degenerate stars, astrometric \citep{1999A&A...348.1040D} and spectroscopic radial velocities agree to better than 30~{\ms} and that the cluster has an internal velocity dispersion of $\simeq 340~${\ms}. 
Therefore, we computed the stars' astrometric velocity ($V_{\rm astro}$) using the Hyades centre and velocities derived from the {\it Gaia} Early Data Release~3 (EDR3). We considered the cluster centre given by \citet{2018MNRAS.477.3197R}, [x,y,z] = [$-44.16 \pm 0.74$, $0.66 \pm 0.39$, $-17.76 \pm 0.41$] pc, and UVW velocities from \cite{2018A&A...616A..10G}: [U, V, W] = [$-6.059 \pm 0.031$, $45.691 \pm 0.069$, $5.544 \pm 0.025$] \kms.

Following \citet{Leao+2019}, we computed $V_{\rm clus}$ accounting for its dependency on the star's right ascension $\alpha$:
\begin{equation}
V_{\rm clus} = (0.034\times \alpha - 2.12)~\kmsmath
\end{equation}
and, for the purpose of this work, it is not relevant whether this dependency is due to the cluster rotation \citep{Leao+2019} or the cluster disruption \citep{2020MNRAS.498.1920O}.

Given the accuracy of {\it Gaia} measurements, the astrometric radial velocities have an associated error of less than 80~\ms, so the cluster velocity dispersion dominates the uncertainty in $V_{\rm r}$. Nonetheless, $V_{\rm clus}$ represents a minor correction, by at most $\simeq 270~$\ms, for our stars.

By assuming that the uncertainty on $V_{\rm r}$ is dominated by the cluster dispersion velocity ($\simeq 340~${\ms}), we exclude the possibility that the observed WDs are affected by peculiar motions. The velocity dispersion of the Hyades has been measured with non-degenerate stars, but WDs may be subject to anisotropic motions. For instance,  \citet{2018MNRAS.480.4884E} show that the presence of asymmetric kicks at a level of $\simeq$0.75 {\kms} reproduces the distribution of WDs in wide binaries. Other authors identified WDs that can be associated with the Hyades, but are escaping from the cluster (see, among others \citealt{2012A&A...547A..99T, 2018MNRAS.477.3197R}), so it can be questioned whether the WDs in our sample can experience anisotropic velocities. Adopting a typical cooling time of 200 Myr \citep{Salaris+2018}, and the conservative assumption that the anisotropic velocity is produced at the beginning of the cooling sequence, a velocity of 0.75 {\kms} would produce a displacement of the star of $\simeq 153\,$pc during the cooling sequence. Of all the WD in our sample, GD~52 is the most distant from the cluster centre, at 15.7 pc. Therefore, we conclude that any anisotropic velocity larger than $\simeq$ 0.08 \kms\ can be safely neglected in our analysis.

Additional motions in the stellar atmospheres ($V_{\rm add}$) includes convective motions. These are dominant in solar-type stars but can be negligible in WD atmospheres as the theory predicts negligible to no convective shift in the NLTE core of the Balmer lines for $\teff \gtrsim 14\,500\,$K  \citep{Tremblay+2011}, which is the case for all stars in our sample (see Sectin~\ref{sec:wd_params}). Finally, any other additional contributions to $V_{\rm add}$ discussed by \citet{Leao+2019} (e.g. stellar rotation and activity) were found to be negligible.

\subsection{Gravitational redshift measurements}\label{sec:gr_espresso}
Velocities have been measured by fitting the H$\alpha$ line with a quadratic function for the wings and a Gaussian for the NLTE core. A second Gaussian was necessary for several spectra to fit an additional narrow component around the line core originating from the residuals of an imperfect sky subtraction. We opted to analyse each spectrum individually (instead of combining the different observations) to obtain realistic estimates of the fitting errors.
$\Delta V$ measurements from the H$\beta$ line suffer from much higher uncertainties because the H$\beta$ NLTE core is shallower and broader than the H$\alpha$ one, and the spectra have lower $S/N$ in the region around H$\beta$. As such, the H$\beta$ and higher Balmer lines recorded in the ESPRESSO spectra were used for cross-checking the H$\alpha$ velocities, not for measuring $\Delta V$.

\begin{figure*}
\centering
{\includegraphics[width=.8\textwidth]{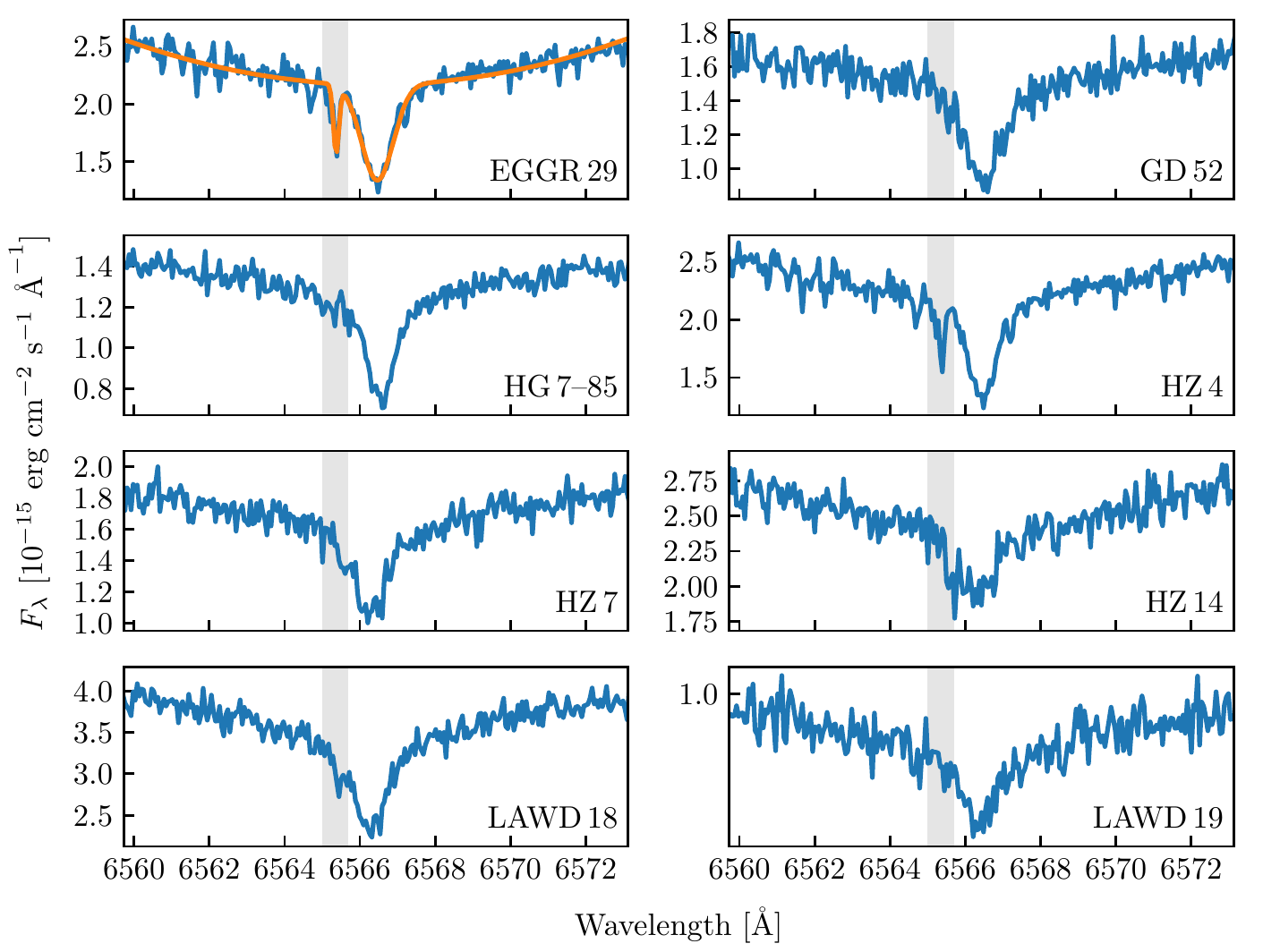}}
\caption{ Close-up on the H$\alpha$ NLTE core of sample ESPRESSO spectra for the eight Hyades WDs. As an example, for the top left spectrum (EGGR\,29), the best-fitting model is overplotted in orange. The photospheric H$\alpha$ is modeled with a  quadratic polynomial and the NLTE core with a Gaussian. In some spectra a second narrow absorption line, whose position is highlighted by the grey band, is visible on the blue side of the core (see e.g. HZ\,4 and EGGR\,29), resulting from the imperfect subtraction of the H$\alpha$ sky emission line. We fitted this residual with a second Gaussian. Without it, the measured $\Delta V$ would have been systematically smaller.}
\label{spectra}
\end{figure*}

Figure~\ref{spectra} shows one spectrum for each star in the region of the H$\alpha$ line. As an example, the best-fitting model is over-plotted in the case of EGGR\,29. Narrow absorption lines resulting from an imperfect sky subtraction are clearly visible in the spectra of EGGR\,29, HZ\,4, HZ\,7 and LAWD\,18. 

For each object, the corresponding $\Delta V$ measurements have been averaged and weighted by the inverse square of the fitting error, to minimise the contribution of inaccurate fit results to the average value. We aimed to reach an uncertainty in the measured $\Delta V$ smaller than or comparable to the cluster velocity dispersion (340 \ms), which is the case for five of the eight WDs. For three objects, namely HZ\,4, LAWD\,18 and HZ\,14, the $\Delta V$ uncertainties lie around 0.5~\kms. Table~\ref{DopplerShifts} reports the average $\Delta V$ for each star, the associated error as computed from the weighted average of the measurements, and the number of spectra used for each average. 

The table also provides the gravitational redshift $V_{\rm z}$ of each star, which we computed using Equation~\ref{eqgr}. Given that we need absolute $V_{\rm z}$ measurements, it is worth reporting that we have adopted vacuum values equal to 6564.6081\,\AA\, and 4862.6829\,\AA\, for H$\alpha$ and H$\beta$, respectively, as derived by \citet{2004ApSpe..58.1469R}.
The errors on $V_{\rm z}$ have been computed by adding the uncertainty given by the cluster velocity dispersion of  340~\ms\ added in quadrature to the uncertainty of the $\Delta V$ measurements. The final errors vary between 0.7 and 1.6 per cent and determine the accuracy of the $M/R$ measurements. We computed the latter using Equation~\ref{eq:mr_eq} and the corresponding values are reported in the last column of Table~\ref{values}. In the following, we refer to these results as $M/R^{\mathrm ESPRESSO}$, to differentiate from the ratios derived from the fit to the \textit{Gaia} photometry (Section~\ref{sec:wd_params}).

\subsection{Systematic uncertainties}
Our velocity measurements agree with those by \citet{Luca+2019} within the 2~\kms uncertainty quoted in that work for the combined UVES and HIRES observations. However, for five of the six stars that have been analysed in this work as well as by \citet{Luca+2019}, the ESPRESSO $\Delta V$s are systematically larger by $\simeq 2~$\kms. The reasons for this systematic difference are not obvious. 
There are, however, several reasons leading us to believe that the ESPRESSO velocities are of superior quality and free of the systematic uncertainties that affected the UVES and HIRES velocities, as detailed below:
\begin{itemize}
\item \citet{Luca+2019} used a literature value equal to 6562.801\,\AA\, in air as reference value for the H$\alpha$ central wavelength. The vacuum value adopted in this work is the center-of-gravity of vacuum wavelengths from recent measurements, which corresponds to 6562.7952\,\AA\, when in air. The difference between the two adopted values accounts for a difference of 265\,\ms\, between our results and those by \citet{Luca+2019}. 

\item UVES and HIRES are not in vacuum, hence all reference wavelengths are reported to wavelengths in air assuming standard air pressure and temperature values. 
Any changes in pressure and/or temperature would cause the air refraction index $n$ to vary by $\Delta n \simeq 1 \times 10^{-6}$/\textdegree Celsius and $\Delta n \simeq 1.3 \times 10^{-6}$/Kpascal \citep{1996ApOpt..35.1566C}. A variation of $\Delta n \simeq 7\times 10^{-6}$ would be sufficient to produce a shift of 2\,\kms around the H$\alpha$ therefore, if the observing conditions were different from the assumed standard values, they would cause (systematic) shifts in the wavelength calibration. Notice that, while for precise velocity determinations only  the difference in atmospheric conditions between the time of the wavelength calibration and the time of the observations matters, for accurate measurements the difference in atmospheric parameters between the time of calibrations and the assumed standard values is relevant. Given the instrument configuration, ESPRESSO observations do not suffer from these uncertainties.

\item For slit instruments such as UVES and HIRES, the observed $\Delta V$ depends on the position of the photo-centre of the star in the slit width, \citep{Luca+2015}. In contrast, ESPRESSO is fibre-fed and the light is scrambled to produce a stable spectrograph injection with a long-term stability of a few \ms. Since the same light path is used by the calibration light, this implies also high accuracy.

\item Finally, UVES and HIRES wavelength calibrations are known to suffer from distortions of several hundreds \ms \citep{Whitmore+2015}. This is not the case for ESPRESSO, as shown using a Laser Frequency Comb \citep{Tobias+2021}. This could also generate systematic differences.
\end{itemize}

As for the analysis, contaminating sky lines may also introduce a systematic effect. ESPRESSO resolving power is much higher ($\simeq 140\,000$) than that of the UVES ($\simeq 18\,500$) and HIRES ($\simeq 40\,000$) observations. In general, this is not so critical for the H$\alpha$ NLTE core, which typically is as broad as $\simeq 1.2\,$\AA. However, the gravitational redshift and the positive radial velocity of the cluster shift the H$\alpha$ absorption line from the WD photosphere always redwards the H$\alpha$ sky emission line. In the case of improper sky subtraction, sky residuals skew the velocity measurements towards lower values. The $\Delta V$ values measured from the spectra with a double Gaussian fit are systematically larger than the measurements from the same spectra if obtained from a single Gaussian fit.
Given that such an approach has never been used in the past, it is possible that the combination of the lower quality (mainly in resolution) of the previous observations, coupled with neglecting the contribution of sky contamination, are an important  cause of the systematic discrepancy between the ESPRESSO observations and the previous (UVES and HIRES) ones.

\begin{table}
\setlength{\tabcolsep}{4pt}
\caption{Velocity measurements for the eight Hyades WDs. Column 2 reports the number of spectra (N) analysed for each object.}
\label{DopplerShifts}      
\centering  
\small
\begin{tabular}{l c c c c c  }        
\hline\hline                 
Star & N & $\Delta V$  & $V_{\rm astro}$ &  $V_{\rm clus}$ & $V_{z}$ \\    
     &   & [\kms]     & [\kms]      &  [\kms]     & [\kms] \\    
\hline
HZ\,14    & 4 &  76.2  $\pm$ 0.5  &  41.26 $\pm$ 0.08   & 0.27   & 34.7 $\pm$ 0.6 \\
LAWD\,19  & 4 &  74.89 $\pm$ 0.05 &  39.64  $\pm$ 0.08 & 0.16   & 35.1 $\pm$ 0.3 \\      
HZ\,7     & 3 &  76.2  $\pm$ 0.2  &  40.50 $\pm$ 0.08   & 0.21   & 35.5 $\pm$ 0.4  \\  
LAWD\,18  & 3 &  75.5  $\pm$ 0.4  &  39.24 $\pm$ 0.08  & 0.12   & 36.2 $\pm$ 0.5 \\
HZ\,4     & 4 &  83.2  $\pm$ 0.5  &  36.38 $\pm$ 0.08  & -0.12  & 46.9 $\pm$ 0.6 \\
EGGR\,29  & 3 &  94.0  $\pm$ 0.3  &  37.63 $\pm$ 0.08  & 0.0    & 56.3 $\pm$ 0.5 \\
HG\,7--85  & 4 &  88.8  $\pm$ 0.2  &  37.16 $\pm$ 0.08  & -0.05  & 51.7 $\pm$ 0.4 \\
GD\,52    & 4 &  86.0  $\pm$ 0.3  &  32.52 $\pm$ 0.08   & -0.15  & 53.6 $\pm$ 0.4 \\
  \hline
\end{tabular}
\end{table}

\subsection{WD parameters}\label{sec:wd_params}
We used the \textit{Gaia} EDR3 data to derive physical parameters, namely effective temperatures, masses, and radii, for the WD in our sample, and compare them with our $M/R$ measurements obtained from the gravitational redshifts. Specifically, we used the \textit{Gaia} EDR3 parallaxes, colours  and magnitudes in combination with the synthetic colours of hydrogen-atmosphere WDs provided by the Montreal  group\footnote{\url{https://www.astro.umontreal.ca/~bergeron/CoolingModels/}} \citep{Bergeron+1995,Holberg+2006,Tremblay+2011}, updated in January 2021, and including the synthetic photometry in the \textit{Gaia} EDR3 $G$, $G_\mathrm{BP}$ and $G_\mathrm{RP}$ filter passbands.

The \textit{Gaia} parallaxes for the eight WDs in our sample have a precision of 0.2\,per cent or better. However, their accuracy is affected by well-known systematics associated with imperfections in the instruments and in the data processing methods \citep{Lindegren+2021}. The mean value of this systematic error, the so-called zero point $\varpi_{\mathrm{ZP}}$, can be modelled according to the \textit{Gaia} magnitudes and colours, as well as the ecliptic latitude of the source. We used the python script \texttt{gaiadr3\_zeropoint}\footnote{\url{https://gitlab.com/icc-ub/public/gaiadr3_zeropoint}} provided by the \textit{Gaia} consortium to correct the parallaxes of our Hyades WDs for their corresponding zero points. These corrections ranged between 0.02 and 0.03~mas. Moreover, as described in \citet{Riello+2021}, we corrected the \textit{Gaia} $G$-band photometry for known systematic effects using the python code \texttt{gaiaedr3-6p-gband-correction}\footnote{\url{https://github.com/agabrown/gaiaedr3-6p-gband-correction}} provided by the \textit{Gaia} collaboration.

We used the \textit{Gaia} parallaxes to compute the absolute $G$, $G_\mathrm{BP}$ and $G_\mathrm{RP}$ magnitudes. First, we estimated the reddening correction for each object following the method from \citet{Gentile+2019,Gentile+2021}, which, given the proximity of the Hyades cluster, results in a negligible reddening of $\lesssim 10^{-3}$~mag.
We then performed a fit to the grid of synthetic magnitudes using the Markov chain Monte Carlo (MCMC) implementation for python, \thisplain, developed by \citet{Foreman-Mackey+2013}. The free parameters of the fit are the WD surface gravity and temperature, for which we assumed flat priors in the ranges covered by the grid of synthetic magnitudes (1500\,K $\leq \teff \leq 150\,000$K and $7.0 \leq \logg \leq 9.0$, with the surface gravity $g$ given in \textit{cgs} units). To account for the uncertainties related to the parallaxes, we assumed a Gaussian prior, centred on the parallax value and weighted by the corresponding parallax uncertainty. Figure~\ref{fig:corner} gives an example of the parameter covariances and posterior distribution.
Our fitting procedure returned the best-fitting \teff\ and \logg\ for which the grid of synthetic magnitudes provides the corresponding WD mass, bolometric magnitude ($M_{\mathrm bol}$), and cooling age.
The combination of surface gravity and the mass, in turn, returns the radius of the WD. 
Following the nomenclature introduced in Section~\ref{sec:gr_espresso}, we dubbed the corresponding ratios as $M/R^{\mathit{Gaia}}$.   

Assuming the IFMR from \citet{Salaris+2009}, we computed the mass of the WD progenitor and, following \citet{Salaris+2018}, the pre-WD lifetime. The latter combined with the WD cooling time gives the total age of each object. Table~\ref{values} reports the results of this fitting procedure. For each star, Table~\ref{values} also reports the \textit{Gaia} absolute magnitude in $G$-band ($M_{\rm G}$), which will be used in the discussion in the following Sections, and, for comparison, the $M/R^{\mathrm ESPRESSO}$ from Section~\ref{sec:gr_espresso}. We discuss in Appendix~\ref{ap:comparison} the differences between these results and those reported by \citet{Luca+2019}.

\begin{table*}
\caption{Best-fit parameters derived from the photometric fit to the \textit{Gaia} data (Section~\ref{sec:wd_params}). The last column reports the $M/R$ derived from the gravitational redshifts obtained from the analysis of the ESPRESSO data (Section~\ref{sec:gr_espresso}).
}       
\label{values}      
\centering  
\setlength{\tabcolsep}{4pt}
\begin{tabular}{l  c c c c c c c c c c}        
\hline\hline                 
      & \multicolumn{8}{c}{\multirow{2}{*}{Photometric fit}} && \multicolumn{1}{c}{Gravitational} \\
          & \multicolumn{8}{c}{} && \multicolumn{1}{c}{redshift} \\          
 \cmidrule{2-9}  \cmidrule{11-11}
Star & \teff                            & \logg             & $R$  & $M$ & $M_{\mathrm bol}$ & $M_{\rm G}$                & Total age & $M/R^{\mathit{Gaia}}$ && $M/R^{\mathrm ESPRESSO}$ \\
            & [K] & & [0.01 R$_\odot$] & [M$_\odot$] & [mag] & [mag] & [Myr] & [M$_\odot$/R$_\odot$] && [M$_\odot$/R$_\odot$]\\

\hline
HZ\,14   & 26\,762 $\pm$ 540 & 8.11 $\pm$ 0.03 & 1.23 $\pm$ 0.02 & 0.703 $\pm$ 0.016 & 7.7 $\pm$ 0.1 & 10.387 & 662 & 57 $\pm$ 2 && 54.5 $\pm $ 0.9\\
LAWD\,19 & 23\,449 $\pm$ 410 & 8.09 $\pm$ 0.02 & 1.23 $\pm$ 0.02 & 0.688 $\pm$ 0.012 & 8.2 $\pm$ 0.1 & 10.635 & 767 & 55.8 $\pm$ 1.8 && 55.1 $\pm$0.5 \\
HZ\,7    & 20\,493 $\pm$ 390 & 8.08 $\pm$ 0.02 & 1.239 $\pm$ 0.018 & 0.674 $\pm$ 0.012 & 8.80 $\pm 0.08$  & 10.867 & 880 & 54.4 $\pm$ 1.7 && 55.8 $\pm$ 0.6 \\
LAWD\,18 & 18\,882 $\pm$ 310 & 8.10 $\pm$ 0.02 & 1.221 $\pm$ 0.017 & 0.682 $\pm$ 0.012 & 9.20 $\pm$ 0.08 & 11.050 & 871 & 55.9 $\pm$ 1.6 && 56.9 $\pm$ 0.8 \\
HZ\,4    & 14\,243 $\pm$ 170 & 8.27 $\pm$ 0.01 & 1.071 $\pm$ 0.010 & 0.781 $\pm$ 0.009 & 10.70 $\pm$ 0.06 & 11.829 & 754 & 72.9 $\pm$ 1.5 && 74 $\pm$ 1 \\
EGGR\,29 & 15\,075 $\pm$ 280 & 8.36 $\pm$ 0.02 & 1.006 $\pm$ 0.016 & 0.835 $\pm$ 0.013 & 10.59 $\pm$ 0.08 & 11.860 & 630 & 83 $\pm$ 3 && 88.4 $\pm$ 0.8 \\
HG\,7--85 & 14\,288 $\pm$ 170 & 8.34 $\pm$ 0.01 & 1.017 $\pm$ 0.010 & 0.825 $\pm$ 0.008 & 10.80 $\pm$ 0.06 & 11.932 & 701 & 81.1 $\pm$ 1.7 && 81.2 $\pm$ 0.6 \\
GD\,52   & 13\,620 $\pm$ 180 & 8.37 $\pm$ 0.01 & 0.993 $\pm$ 0.009 & 0.842 $\pm$ 0.008 & 11.05 $\pm$ 0.06 & 12.051 & 766 & 84.7 $\pm$ 1.7 && 84.2 $\pm$ 0.6 \\
 \hline
\end{tabular}
\end{table*}

\section{Discussion }
\label{summary}
The results in Table~\ref{values} show an excellent agreement between $M/R^{\mathit{Gaia}}$ and $M/R^{\mathrm ESPRESSO}$.
The largest fractional difference ($(M/R^{\mathrm ESPRESSO}-M/R^{\mathit{Gaia}})/(M/R^{\mathrm ESPRESSO})\simeq 6$\,per cent) is observed for EGGR\,29, which is likely a special case: as discussed below, this star possibly descends from a blue straggler.  

After excluding EGGR\,29, the average fractional difference between the $M/R$ evaluated from gravitational redshifts and photometry is $-0.8\,$per cent, with a dispersion of 2.2\,per cent. 
This agreement is remarkable, since the two methods employed to estimate $M/R^{\mathit{Gaia}}$ and $M/R^{\mathrm ESPRESSO}$ are entirely independent, as the gravitational redshift method does not use any information from the theoretical evolutionary tracks. We also anticipate  that such a good agreement between models and observations is not valid for spectroscopically determined WD parameters, as discussed in Section~\ref{sec:par_spec}.
  
\subsection{A consistent picture}\label{sec:discussion}
In our analysis, we did not consider a unique characteristic of the sample: the  WDs  belong to a cluster, so they are all expected to have the same total (progenitor plus cooling) age. However, it is clear from Table~\ref{values} that the ages derived from the photometric fit (Section~\ref{sec:wd_params}) are characterised by a large spread, considering that the typical errors on the total ages are by less than ten per cent. One important point to consider is that the total ages depend critically on the IFMR adopted for their computation. The IFMR given by \citet{Salaris+2009} was calculated as an average value from many open clusters, under the assumption that all follow the same IFMR. This practice is very common, and the detailed shape of the general IFMR is discussed in several works \citep[see also][]{Cummings+2018}.

A more detailed approach to derive the IFMR specific for the cluster in consideration is to use the main-sequence turnoff. If this is well defined in the cluster CMD, then it is possible to derive the main sequence turnoff age and thus, the age of the cluster. Following this approach, \citet{2018A&A...616A..10G} obtained an age of 790\,Myr for the Hyades. \citet{Salaris+2018} then adopted this value in their study of the cluster's WDs and derived an updated IFMR for the Hyades. As discussed by \citet{Salaris+2019}, this IFMR turned out to be slightly different from that determined by previous studies \citep{Salaris+2009,Cummings+2018}. In particular, the IFRM by \citet{Cummings+2018} systematically overestimates by $\simeq 0.02\,\mathrm{M_\odot}$ the WD final masses for progenitors in the mass range $2.5-3.5\,\mathrm{M_\odot}$.

To account for these results, we have computed new isochrones for the Hyades cluster. We adopted the WD cooling tracks and progenitor lifetimes from \citet{Salaris+2018} since the WD masses and cooling times derived by these authors are consistent at the level of better than one and nine per cent, respectively, with those determined from our photometric analysis. Assuming a cluster age of 790\,My \citep{2018A&A...616A..10G}, we built an IFMR that, in the initial mass range $2.5-3.5\,\mathrm{M_\odot}$, returns final masses that are $0.02\,M_{\odot}$ smaller compared to those returned by the IFMR from \citet{Cummings+2018}. Outside this initial mass range, our IFMR smoothly converges to the original IFMR of \citet{Cummings+2018}.

\begin{figure}
\centering{\includegraphics[width=\columnwidth]{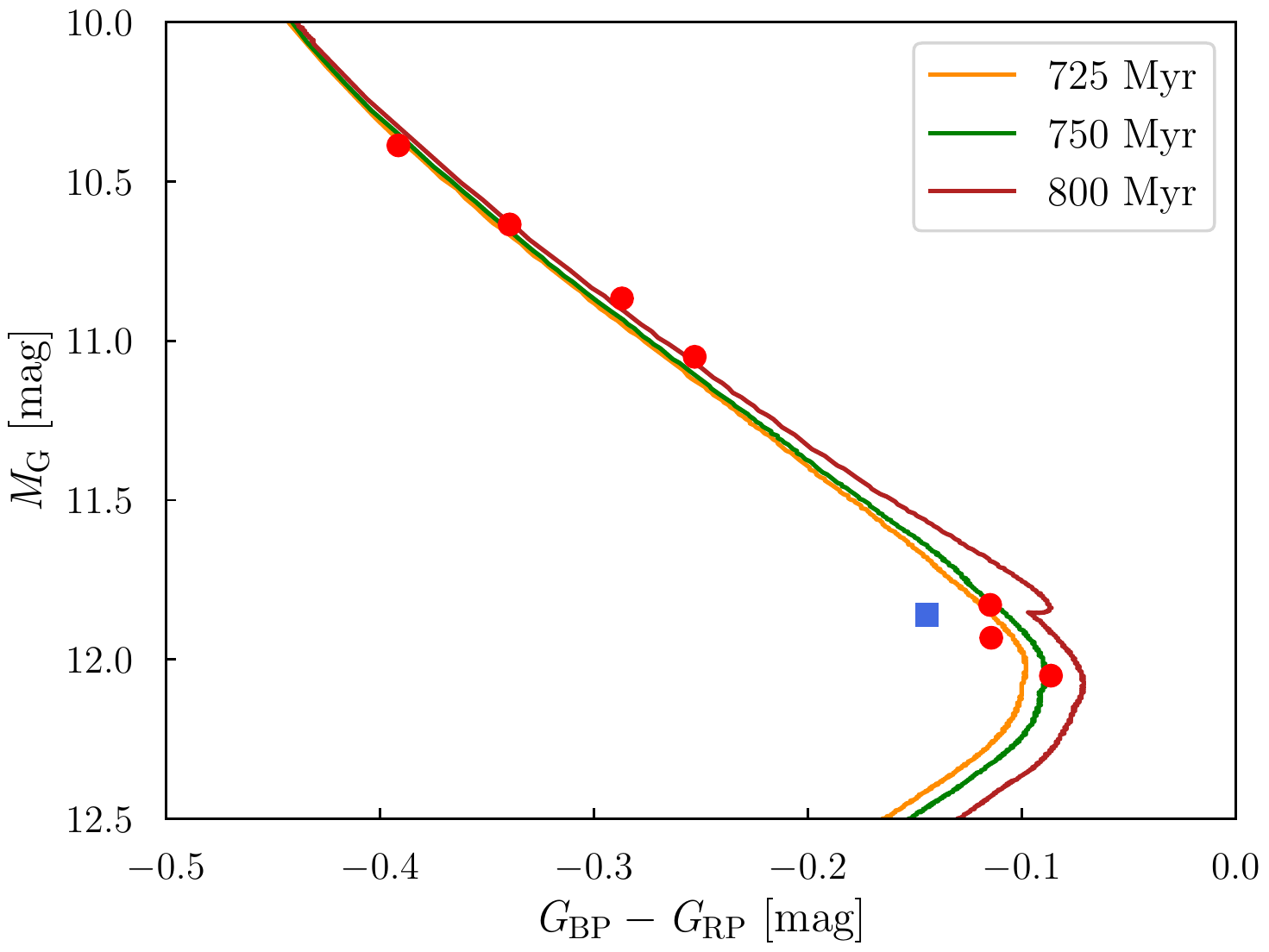}}
\caption{\textit{Gaia} CMD of the eight Hyades WDs. The 725, 750 and 800\,Myr isochrones are displayed. They have been calculated employing the models from \citet{Salaris+2018} and an updated IFMR (see the text for more details). It is possible to confine all stars but EGGR\,29 (blue square) in this age range.}
\label{cmd2}
\end{figure}

\begin{figure}
\centering{\includegraphics[width=\columnwidth]{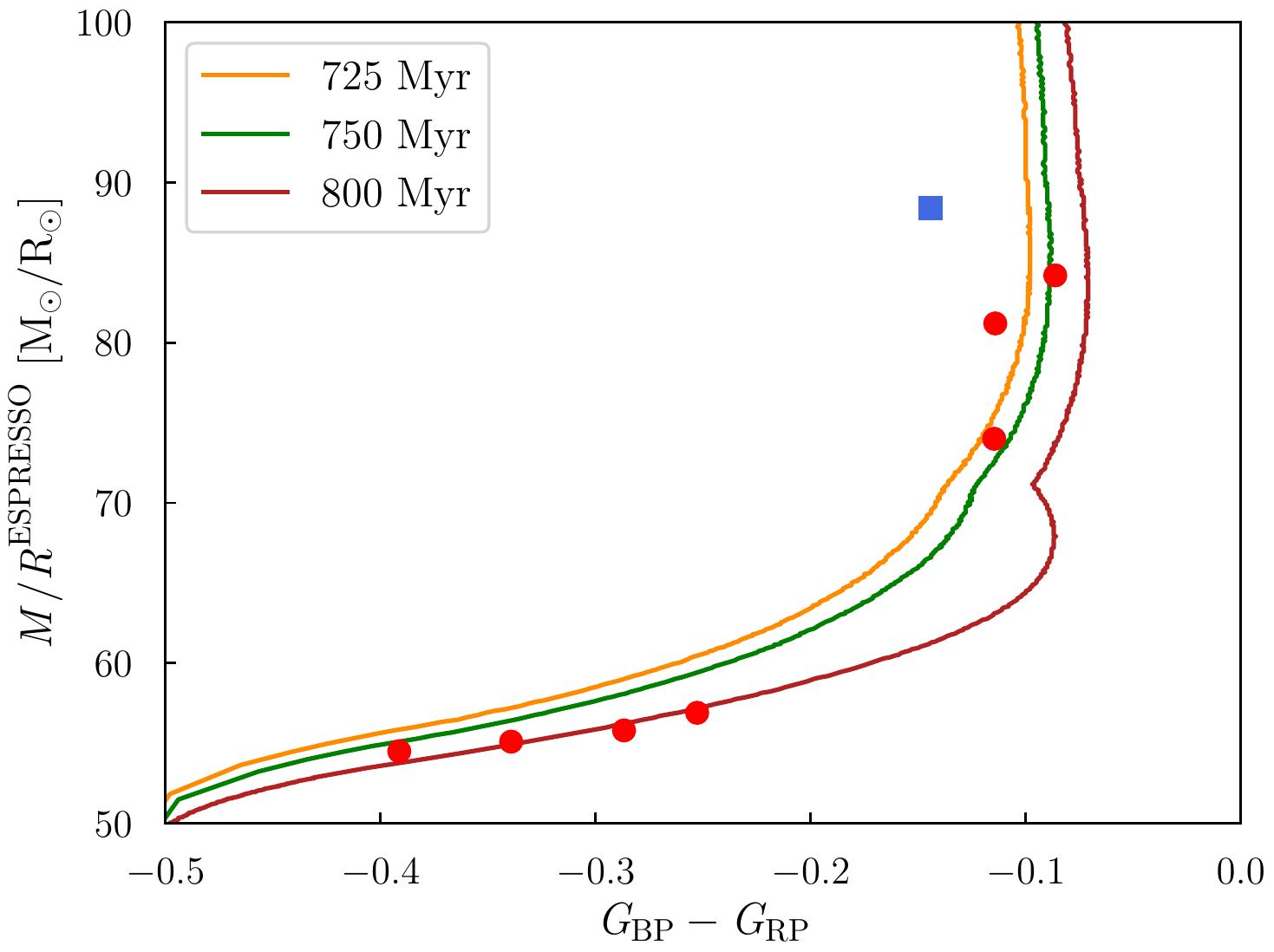}}
\caption{$M/R$ vs colour diagram for the eight Hyades WDs, where the  same isochrones of Figure~\ref{cmd2} are included. All stars but EGGR~29 (blue square) fit in the 725-800 Myr age range. The uncertainties are smaller than the sizes of the points.}
\label{msursec}
\end{figure}

The isochrones with ages between 725 and 800 Myr are those that better reproduce the observed CMD of the Hyades WDs in our sample, as shown in Figure~\ref{cmd2}. This is  expected, given that this IFMR was built assuming a cluster age of 790\,Myr. What it is striking is that the WD parameters derived from these isochrones are also capable to reproduce the measured $M/R$, as shown in Figure~\ref{msursec}.

We therefore obtain a consistent picture of the Hyades WDs by using the \citet{Salaris+2018} isochrones, which are capable to reproduce the \textit{Gaia} magnitudes and colours, as well as for the gravitational redshift-based $M/R$, with all stars (except EGGR\,29) obeying a single IFMR and having ages in the range $725-800\,$Myr.

\subsection{Hyades WD parameters from spectroscopy}\label{sec:par_spec}
It is possible to derive masses and radii of WDs from spectroscopy by determining stellar effective temperatures and gravities from the modelling of the Balmer lines in combination with a theoretical mass-radius relationship.
Several authors have derived \teff\ and \logg\, for Hyades WDs in the past 20 years \citep{2001ApJ...563..987C, 2009A&A...505..441K, 2010ApJ...714.1037L, 2011ApJ...743..138G, Cummings+2018}. In particular, \citet{Cummings+2018} employed updated analysis techniques (that better account for the different sources of noise in the fitting procedure) and models (that account for improved Stark broadening calculation) and therefore we limit our comparison to their results.  

These are summarized in Table~\ref{temperature}, where we also computed the corresponding $M/R$ values from the given $\logg$ and mass. The related errors have been computed with a Monte Carlo calculation. For ease of comparison, our $M/R^{\mathit Gaia}$ and $M/R^{\mathrm ESPRESSO}$ are repeated in this Table.

\begin{table*}
\caption{Spectroscopic stellar parameters for the Hyades WDs from \protect{\citet{Cummings+2018}} and \protect{\citet{Gentile+2021}} in comparison with our results from the fit to the \textit{Gaia} photometry, and with those from our gravitational redshift measurements.}
\label{temperature}      
\centering  
\setlength{\tabcolsep}{4pt}
\begin{tabular}{l c c c c c c c c c c c c}        
\hline\hline                 
     & \multicolumn{4}{c}{Spectroscopic fit} && \multicolumn{4}{c}{Photometric fit} \\
     & \multicolumn{4}{c}{\citep{Cummings+2018}} && \multicolumn{4}{c}{\citep{Gentile+2021}}  && \multicolumn{2}{c}{This work}\\\cmidrule{2-5}\cmidrule{7-10}\cmidrule{12-13}  
Star & 	\teff	& \logg & $M$ & $M/R$ && \teff & \logg & $M$ & $M/R$ && $M/R^{\mathit{Gaia}}$ & $M/R^{\mathrm ESPRESSO}$\\
      & [K]     &       & [M$_\odot$] & [M$_\odot$/R$_\odot$] && [K]     &       & [M$_\odot$] & [M$_\odot$/R$_\odot$] && [M$_\odot$/R$_\odot$]  & [M$_\odot$/R$_\odot$] \\
            
\hline
HZ\,14    & 27\,540(400) & 8.15(5) & 0.726(3) & 61(4) && 27\,557(570) & 8.14(3) & 0.721(18) & 60(2) && 57 (2) & 54.5(9)\\
LAWD\,19  & 25\,130(380) & 8.12(5) & 0.704(29) & 58(3) && 24\,172(490) & 8.12(3) & 0.703(16) & 58.3(1.9) && 55.8(1.8) & 55.1(5)\\
HZ\,7     & 21\,890(350) & 8.11(5) & 0.691(3) & 60(3) && 21\,152(435) & 8.11(3) & 0.689(15) & 56.7(1.8) && 54.4(1.7) & 55.8(6) \\
LAWD\,18  & 20\,010(320) & 8.13(5) & 0.700(3) & 59(3) && 19\,410(400) & 8.12(3) & 0.695(15) & 58.1(1.8) && 55.9(1.6) & 56.9(8)\\
HZ\,4     & 14\,670(380) & 8.30(5) & 0.797(32) & 76(5) && 14\,516(205) & 8.293(17) & 0.792(11) & 75.3(1.6) && 72.9(1.5) & 74(1)\\
EGGR\,29  & 15\,810(290) & 8.38(5) & 0.850(32) & 86(5) && 15\,425(280) & 8.37(2) & 0.844(15) & 85(2) && 83(3) & 88.4(8) \\
HG\,7--85 & 14\,620(60)  & 8.25(1) & 0.765(6) & 70.4(9) && 14\,669(210) & 8.361(16) & 0.837(11) & 83.7(1.6) && 81.1(1.7) & 81.2(6) \\
GD\,52    & 14\,820(350) & 8.31(5) & 0.804(32) & 77(5) && 14\,008(220) & 8.389(15) & 0.85(1) & 87.3(1.65) && 84.7(1.7) & 84.2(6) \\
 \hline
\end{tabular}
\end{table*}

Even if the effective temperatures derived from the spectroscopic and photometric fits are in agreement within the nominal uncertainties, it is clear that the spectroscopic values are systematically higher than those obtained from the photometric fits. Moreover, the surface gravities and masses obtained from the spectroscopic analysis are higher than the photometric results for six stars, and lower for two, cool objects. The combination of higher gravities and effective temperatures implies that smaller radii are expected from the spectroscopic method with respect to the photometric one.
This discrepancy has been described in the past in great detail by several authors \citep[see e.g.][]{2019MNRAS.482.5222T, 2019ApJ...871..169G, 2019ApJ...876...67B}. 

As a consequence, the $M/R$ derived from the WD parameters determined by \citet{Cummings+2018} do not agree  with our results, i.e. they differ more than 3$\sigma$ from our model-independent ratios derived from the gravitational redshift ($M/R^{\mathrm ESPRESSO}$).

\begin{figure}
\centering{\includegraphics[width=\columnwidth]{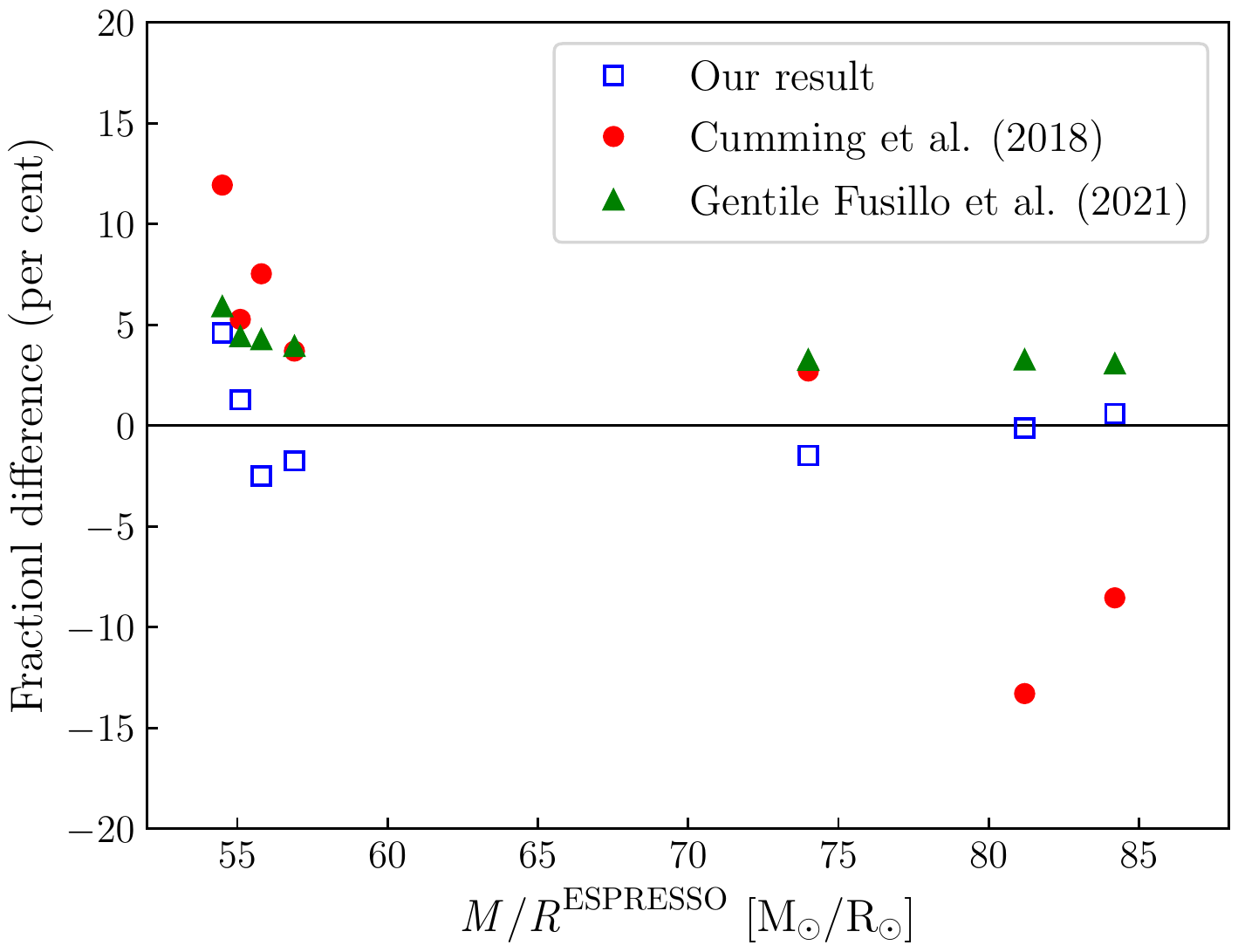}}
\caption{Fractional difference between our $M/R^{\mathrm ESPRESSO}$ and the $M/R$ derived from the WD parameters measured by \citet{Cummings+2018} (red circles), and our $M/R^{\mathrm Gaia}$ and the $M/R$ derived from the WD parameters measured by \citet{Gentile+2021} (green triangles). For comparison, also the difference between $M/R^{\mathrm ESPRESSO}$ and $M/R^{\mathit Gaia}$ is shown (blue squares). EGGR\,29 is not included in the plot because of its peculiar nature (see Section~\ref{sec:eggr29}).}
\label{comparison}
\end{figure}

We find that the fractional difference (computed as ($M/R-M/R^{\mathrm ESPRESSO}$)/$M/R^{\mathrm ESPRESSO}$) between the two datasets\footnote{In this discussion, we do not consider EGGR\,29 because of its peculiar nature (see Section~\ref{sec:eggr29}).}  (Figure~\ref{comparison}) shows a possible trend with $M/R$, with a difference as high as $10-15\,$per cent at the limits of the range covered by our $M/R^{\mathrm ESPRESSO}$ measurements. The observed trend seems to suggest that the spectroscopic results suffer from some unknown systematics. 

Spectroscopic measurements are generally considered the most precise. However, several works argue that the Balmer line profile in WDs needs to be fully understood, either using experimental data \citep{Schaeuble+2019} or new theoretical ingredients \citep{Cho+2022}.
The model-independent $M/R^{\mathrm ESPRESSO}$ measurements from the gravitational redshift provide a new powerful constraint, and, our photometric $M/R^{\mathit Gaia}$ agree better with them than the spectroscopic ones available in the literature for the Hyades WDs. 

It is finally interesting to compare our $M/R^{\mathrm Gaia}$ measurements with the results from other photometric analyses, such as the work by \citet{Gentile+2021}, which used the data from \textit{Gaia} EDR3 to identify and characterise more than 300\,000 WDs in the Milky Way. The difference between our $M/R^{\mathit Gaia}$ and the $M/R$ derived from the WD parameters estimated by \citet{Gentile+2021} is also displayed in Figure~\ref{comparison} as green triangles. On average, the latter are systematically higher than our results by $\simeq 4$~per cent. This difference can be explained by a different photometric zero point for Vega spectra used in the calculation of the WD models used in this work and those used by \citet{Gentile+2021} (N. Gentile Fusillo, private communication).

\subsection{EGGR\,29}\label{sec:eggr29}
\begin{figure}
\centering{\includegraphics[width=\columnwidth]{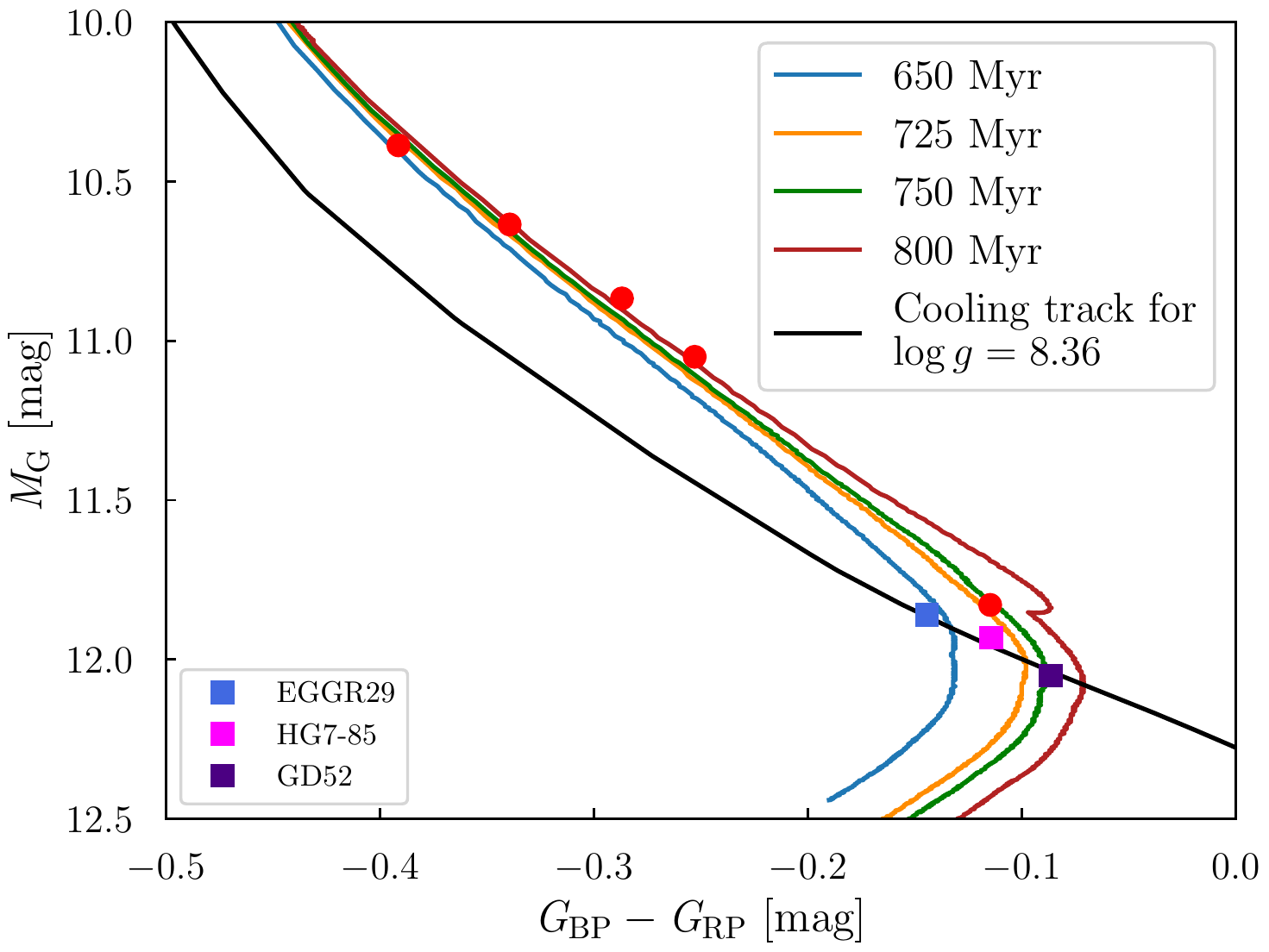}}
\caption{\textit{Gaia} CMD of the eight Hyades WDs. The 650, 725, 750 and 800\,Myr isochrones are displayed (see Section~\ref{sec:discussion}). The three more massive WDs in our sample, EGGR\,29, HG\,7-85 and GD\,52, have similar masses ($M \simeq 0.835\,\mathrm{M_\odot}$) and all sit on the cooling track for $\log g \simeq 8.36$ (Table~\ref{values}) but the position of EGGR\,29 indicates that it is a younger WD, formed $\simeq 150\,$Myr later compared to the other WDs in the Hyades cluster.}
\label{eggr29_cooling}
\end{figure}

In the CMD of Figure~\ref{cmd2}, it is  clear that the star EGGR\,29 does not lie on the same sequence as the other WDs in our sample. Its parallax, proper motions, and astrometric radial velocity from  \textit{Gaia} show that this star is a bona fide member of the Hyades cluster, and its membership has never been questioned in the literature.
However, when compared with the other Hyades WDs of comparable absolute magnitude and masses (HG\,7-85 and GD\,52, $M \simeq 0.835\,\mathrm{M_\odot}$, Table~\ref{values}), it is clear that EGGR\,29 occupy a different position on the cooling track describing the evolution of these stars, indicating that it is characterised by a smaller cooling age. This reflects the fact that EGGR\,29 is the hottest WDs among these three and, therefore, it must have formed as a WD at a later phase than HG\,7-85 and GD\,52 (Figure~\ref{eggr29_cooling}).

Our $M/R$ measurements also show, consistently with inferences from the CMD, that the $M/R$ of EGGR\,29 is $\simeq 3\,$\kms\, larger than those of HG\,7-85 and GD\,52 (Figure~\ref{msursec}), a value that exceeds the measurement uncertainty by more than 7$\sigma$. The magnitude, colours and $M/R$ of EGGR\,29 could be reproduced using isochrones computed assuming an age $\simeq 150\,$Myr younger than that of the other WDs in our sample, as can be seen in Figure~\ref{eggr29_cooling} where an isochrone for 650\,My is displayed.

A WD outlier with similar characteristics, LB~5893, has been observed in the Hyades twin cluster Praesepe \citep{Salaris+2019}. 
The anomalous status of LB~5893 was discovered by \citet{Casewell+2009} from its peculiar position in the initial mass–final mass space. Similarly to EGGR\,29, also for LB\,5893 no other peculiar characteristics are observed. \citet{Casewell+2009} discuss two possible scenarios to explain the peculiarity of LB~5893: either it descends from a blue straggler, or it is the result of differential mass loss during the asymptotic giant branch (AGB) phase. The authors favoured the blue straggler progenitor hypothesis since this WD is the only object in the cluster showing such peculiar properties, while it would be expected that differential mass loss would produce other WD outliers.

Differential mass loss could be a possible explanation for the anomalies observed in EGGR\,29. However, also in this case (as for Praesepe), it would be expected that other deviant objects should be observed among the WDs in the cluster and not just one exception. Therefore, the blue straggler progenitor hypothesis seems to be the most likely also in the case of EGGR\,29. The age of this WD, as derived from the isochrone fitting, is $\simeq 650\,$Myr and the merging that led to formation of the blue straggler progenitor should have occurred on the main sequence, because the typical time needed by two WDs to merge is comparable to or longer than the age of the cluster \citep{Cheng+2020}.

\section{Conclusions} 
We have obtained accurate measurements of $M/R$ for eight bona fide Hyades WDs from ESPRESSO velocity shifts and \textit{Gaia} photometry. A comparison between our gravitational redshift-based $M/R$ measurements and the photometric-based results shows an agreement to better than one\,per cent. This result is quite remarkable, given that the two methods are completely independent and rely on different theories (general relativity in the case of the gravitational redshift, and quantum mechanics in the case of the photometric fit with synthetic WD models).

By using the isochrones computed by \citet{Salaris+2018} and tuned for the Hyades IFMR (assuming the cluster age from the main sequence turn-off fitting), we found a consistent picture that can reproduce both the WD cooling sequences and the $M/R$ observed values, with the age of the cluster being comprised in the narrow range 725-800\,Myr.

One star, EGGR~29, does not populate the same locus of the other Hyades WDs in the CMD and $M/R$-colour diagram but appear to be $\simeq 150\,$Myr younger. EGGR\,29 resembles the peculiar WD LB\,5893 observed in the Hyades-twin cluster Praesepe. As for LB\,5893, we suggest that EGGR\,29 is the remnant of a blue straggler.

Finally, we confirm the presence of a discrepancy between the WD parameters derived from the spectroscopic and from the photometric analysis. This discrepancy is systematic and well-known in the literature. In this context, our accurate $M/R$ measurements add a new constraint that can be used to further refine the currently available synthetic atmosphere models.
 
\section*{Acknowledgements}
The authors thank an anonymous referee for a very careful reading of the manuscript and many suggestions that have greatly improved the quality of the  work. LPA acknowledges the scientific hospitality of the Arcetri Observatory. J.R.M. and I.C.L. acknowledge financial support from the CNPq brazilian agency. The authors thank Nicola Gentile Fusillo for very useful discussions. 
MS acknowledges support from The Science and Technology Facilities Council Consolidated Grant ST/V00087X/1.
This research has made use  of data from the European Space Agency (ESA) mission {\it Gaia} (\url{https://www.cosmos.esa.int/gaia}), processed by the {\it Gaia} Data Processing and Analysis Consortium (DPAC, \url{https://www.cosmos.esa.int/web/gaia/dpac/consortium}). Funding for the DPAC has been provided by national institutions, in particular the institutions participating in the {\it Gaia} Multilateral Agreement.

\section*{Data Availability}
The ESPRESSO spectra are publicly available in the ESO archive.



\bibliographystyle{mnras}
\bibliography{references} 



\appendix
\section{Covariances and posterior distributions}\label{ap:covariances}
\begin{figure}
\centering
{\includegraphics[width=.5\textwidth]{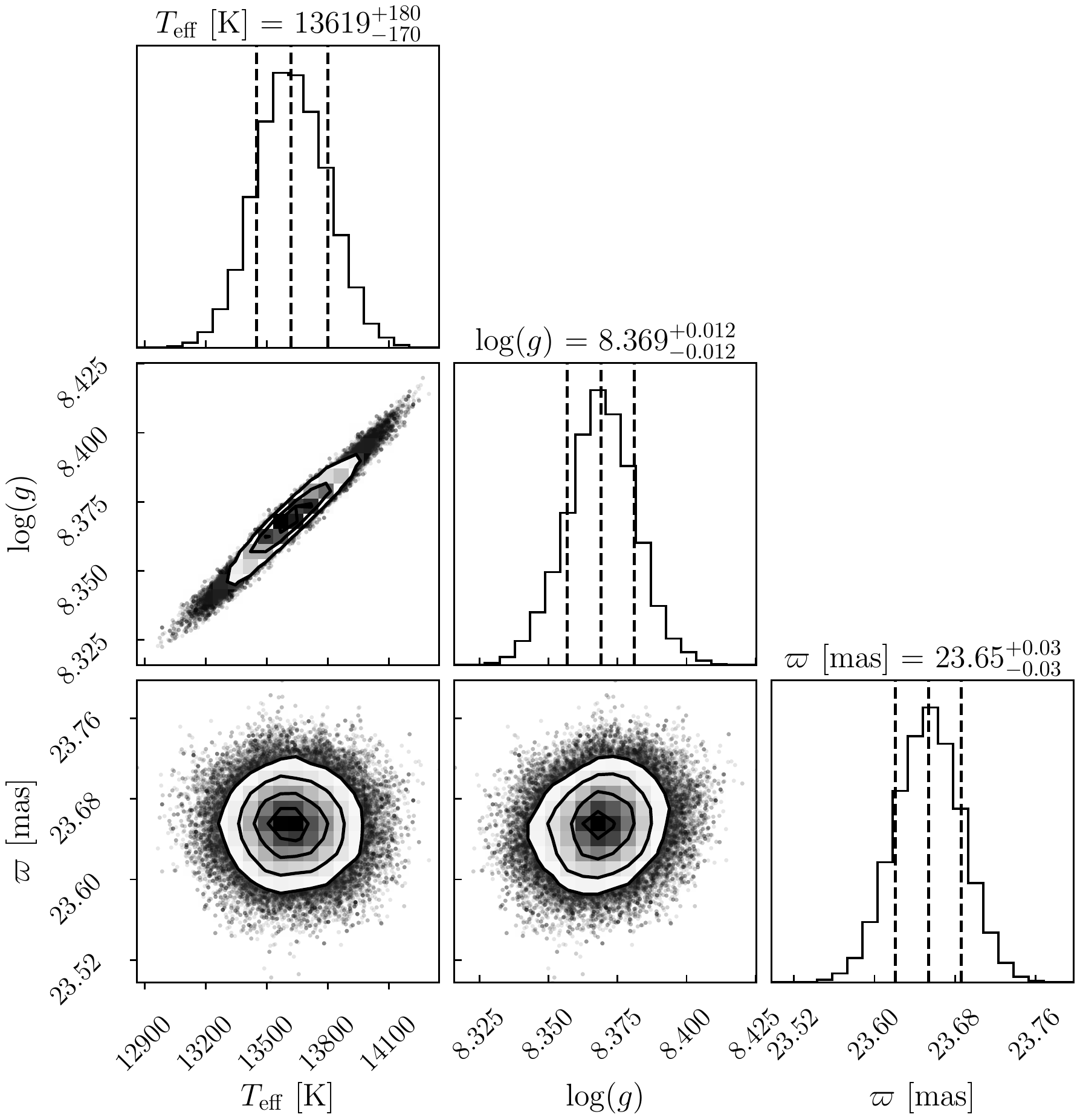}}
\caption{Sample corner plot showing the parameter covariances and posterior distributions for the MCMC fit to the \textit{Gaia} EDR3 photometry for GD52. The 16th, 50th, and 84th quantiles are shown as vertical lines. The figure has been generated using an adaptation of the corner.py package developed by Dan Foreman-Mackey and contributors \citep{corner}.}
\label{fig:corner}
\end{figure}

Figure~\ref{fig:corner} shows a sample corner plot for the MCMC fitting of the star GD\,52.

\section{Comparison with Pasquini et al. (2019)}\label{ap:comparison}
Six (HZ\,4, HZ\,7, LAWD\,18, LAWD\,19, EGGR\,29, and HG\,7--85) out of the eight Hyades WDs considered in this work have also been analysed by \citet{Luca+2019}.
Those authors obtained radii and masses of these objects by performing a fit of their absolute $G$, $G_\mathrm{BP}$ and $G_\mathrm{RP}$ magnitudes, derived using the parallaxes provided by \textit{Gaia} DR2, to the synthetic magnitudes computed by Bergeron and collaborators in 2018 for the corresponding filter passbands.

The masses from this work are, on average, $\simeq 0.03\,M_\mathrm{\odot}$ larger than those reported in Table~2 by \citet{Luca+2019}.
To investigate whether this difference could be related to the different fitting routines employed (least square minimisation routine from \citealt{Luca+2019} vs our MCMC fitting, including a prior on the \textit{Gaia} parallax), we fitted the \textit{Gaia} DR2 data following the procedure described in Section~\ref{sec:wd_params}. The masses thus derived are in good agreement with those reported by \citet{Luca+2019} (with a typical difference by less $\simeq 0.004\,M_\mathrm{\odot}$). We therefore conclude that the difference observed with the masses from \citet{Luca+2019} is not due to the different fitting methods but rather to (i) the improved accuracy of the \textit{Gaia} EDR3 parallaxes (which we also corrected for the zero points), and (ii) the new updated theoretical models that we have employed in the analysis carried out in this work.

\bsp	
\label{lastpage}
\end{document}